\documentclass{article}
\usepackage{ijcai15}

\usepackage{times}

\usepackage{amsmath}
\usepackage{amssymb,bm}
\usepackage{booktabs}
\usepackage{breqn}
\usepackage{caption}
\usepackage{float}
\usepackage{footmisc}
\usepackage{graphicx}
\usepackage{lipsum}
\usepackage{makecell}
\usepackage{multirow}
\usepackage{pbox}
\usepackage{subfigure}
\usepackage{times}
\usepackage{tabu}
\usepackage{tabularx}
\usepackage{verbatim}
\usepackage{url}

\newcommand{\norm}[1]{\ensuremath{\lVert{#1}\rVert}}
\DeclareMathOperator*{\argmin}{arg\,min}

\title{A Synthetic Approach for Recommendation: Combining Ratings, Social Relations, and Reviews}

\author{Guang-Neng Hu$^1$,  Xin-Yu Dai$^1$, Yunya Song$^2$, Shu-Jian Huang$^1$, Jia-Jun Chen$^1$ \\
$^1$National Key Laboratory for Novel Software Technology; $^2$Department of Journalism\\
$^1$Nanjing University, Nanjing 210023, China; $^2$Hong Kong Baptist University, Hong Kong \\
\{hugn, huangsj\}@nlp.nju.edu.cn, \{daixinyu, chenjj\}@nju.edu.cn, yunyasong@hkbu.edu.hk
}

\begin{document}
\maketitle

\begin{abstract}
  Recommender systems (RSs) provide an effective way of alleviating the information overload problem by selecting personalized choices. Online social networks and user-generated content provide diverse sources for recommendation beyond ratings, which present opportunities as well as challenges for traditional RSs. Although  {\em social matrix factorization} (Social MF) can integrate ratings with social relations and {\em topic matrix factorization} can integrate ratings with item reviews, both of them ignore some useful information. In this paper, we investigate the effective data fusion by combining the two approaches, in two steps. First, we extend Social MF to exploit the graph structure of neighbors. Second, we propose a novel framework {\mbox{MR3}} to jointly model these three types of information effectively for rating prediction by aligning latent factors and hidden topics. We achieve more accurate rating prediction on two real-life datasets. Furthermore, we measure the contribution of each data source to the proposed framework.
\end{abstract}

\section{Introduction}

For all the benefits of the information abundance and communication technology, the ``information overload'' is one of the digital-age dilemmas we are confronted with. Recommender systems (RSs) are instrumental in tackling this problem as they help determine which information to offer to individual consumers and allow users to quickly find the personalized information that fits their needs ~\cite{CF92,linden03:amazon,koren09:MF}. RSs are nowadays ubiquitous in various domains and e-commerce platforms, such as recommendation of books at Amazon, musics at Last.fm, movies at Netflix and references at CiteULike.

Social networking and knowledge sharing sites like Twitter and Epinions are popular platforms for users to connect to each other, to participate in online activities, and to generate shared opinions. Social relations and item contents provide independent and diverse sources for recommendation beyond explicit rating information~\cite{ganu09:review,HFT,SoRec,LOCABAL}, which present both opportunities and challenges for traditional RSs.

{\em Collaborative filtering} (CF) approaches are extensively investigated in research community and widely used in industry. They are based on the naive intuition that if \mbox{users} rated items similarly in the past, then they are likely to rate other items similarly in the future~\cite{CF92,sarwar01:itemCF}. Latent factors CF, which learns a latent vector of preferences for each user and a latent vector of attributes for each item, gains popularity and becomes the standard model for recommender due to its accuracy and scalability~\cite{CFSVD98,koren09:MF}. CF models, however, suffer from data sparsity and the imbalance of ratings; they perform poorly on cold users and cold items for which there are no or few data.

To overcome these weaknesses, additional sources of information are integrated into RSs. One research thread, which we call {\em social matrix factorization} (Social MF), is to combine ratings with social relations~\cite{SoRec,SoReg,tranMF,LOCABAL,TrustSVD}. Extensive studies have found higher likelihood of establishing social ties among people having similar characteristics, namely the theory of homophily~\cite{homophily,socialmedia}. Given that interpersonal similarity and effective communication condition, homophilous ties become effective means of social influence~\cite{influence,localInfluence}. Social MF methods factorize rating matrix and social matrix simultaneously.

Another research thread, which we call {\em topic matrix factorization} (Topic MF), is to integrate ratings with item contents or reviews text~\cite{CTR,RMR}. Reviews justify the rating of a user, and ratings are associated with item attributes hidden in reviews~\cite{jakob09review,ganu09:review}. Topic MF methods combine latent factors in ratings with latent topics in item reviews~\cite{HFT,TopicMF}. Nevertheless, both Social MF and Topic MF ignore some useful information, either item reviews or social relations.

There is a tendency towards hybrid methods~\cite{CFCBFdemograph99,CTRSoRec,CCTRSoRec}. These methods all consider diverse sources for recommendation, however, the first two methods are belonging to one-class CF~\cite{OneCF} and hence the dimensions discovered are not necessarily correlated with rating; while the last two methods adopt two components which are not effective~\cite{HFT,LOCABAL}. Hence, it is still a challenge to find an effective way to integrate multiple data sources for recommendation.

In this paper, we investigate the effectiveness of fusing social relations and review texts to rating prediction in a novel way, inspired by the complementarity of the two independent sources for recommendation. The core idea is the alignment between latent factors found by Social MF and topics found by Topic MF. Our main contributions are outlined as follows.
\begin{itemize}
\item {Providing a principled way to exploit ratings and social relations tightly for recommendation, where the tightness means exploiting the graph structure of neighbors;}
\item {Proposing an effective framework MR3 to jointly model ratings, the social network, and item reviews for rating prediction, where the effectiveness means adopting two effective components in some sense;}
\item {Evaluating the proposed model extensively on two real-world datasets to understand its performance.}
\end{itemize}

The organization of this paper is as follows. Problem setting and notations are given in Section \ref{paper:Setting}. In Section \ref{paper:MR3}, we present the two components and details of the proposed framework. In Section \ref{paper:Exp}, we give empirical results on real-life datasets. Concluding remarks with a discussion of some future work are in the final section.

\section{Problem Statement and Notation}\label{paper:Setting}
Suppose there are $I$ users $\mathcal{U}=\{u_1,...,u_I\}$ and $J$ items $\mathcal{V}=\{v_1,...,v_J\}$. Let $R \in \mathbb{R}^{I \times J}$ denote the rating matrix, where $R_{i,j}$ is the rating of user $i$ on item $j$, and we mark a zero if it is unknown. The task of rating prediction is to predict missing ratings from the observed data. Latent factors CF methods like {\em probabilistic matrix factorization} (PMF)~\cite{PMF} exploit ratings for recommender.

Users connect to others in a social network. We use $T \in \mathbb{R}^{I \times I}$ to indicate the user-user social relations; $T_{i,k}$ = 1 if user $i$ has a relation to user $k$ or zero otherwise. Social MF methods like {\em social recommendation} (SoRec)~\cite{SoRec} and {\em local and global} (LOCABAL)~\cite{LOCABAL} integrate social relations for recommender.

Items have content information, e.g., reviews commented by users. The observed data $d_{i,j}$ is the review of item $j$ written by user $i$, often along with a rating score $R_{i,j}$. Topic MF methods like {\em collaborative topic regression} (CTR)~\cite{CTR} and {\em hidden factors and topics} (HFT)~\cite{HFT} integrate item content for recommender.

Both Social MF and Topic MF ignore some useful data sources, either item reviews or social relations. Notations used in this paper are described in Table~\ref{table:notation}.
\begin{table}
\centering
\begin{tabular}{l l}
 \Xhline{2\arrayrulewidth}
 Symbols & Meanings \\
 \hline
 $F$        & dimensionality of latent factors/topics \\
 $R_{i,j}$  & rating of item $j$ by user $i$  \\
 $U_i$      & $F$-dimensional features for user $i$\\
 $V_j$      & $F$-dimensional features for item $j$\\
 $W_{i,j}$  & weight on the rating of item $j$ given by user $i$\\
 $T_{i,k}$  & social relation between user $i$ and $k$ \\
 $C_{i,k}$  & social strength between user $i$ and $k$ \\
 $S_{i,k}$  & social rating similarity between user $i$ and  $k$ \\
 $H$        & $F \times F$-dimensional social correlation matrix\\
 $d_{i,j}$  & review (`document') of item $j$ by user $i$\\
 $w_{d,n}$; $z_{d,n}$ & the $n^{\mathrm{th}}$ word in doc $d$; corresponding topic\\
 $\theta_j$ & $F$-dimensional topic distribution for item $j$ \\
 $\phi_f$   & word distribution for topic $f$ \\
 \Xhline{2\arrayrulewidth}
\end{tabular}
\caption{Notations}
\label{table:notation}
\end{table}

\section{The Proposed Framework}\label{paper:MR3}

\subsection{Matrix Factorization: A Basic Model}
Rating scores are the explicit user feedback and matrix factorization (MF) is a state-of-the-art recommender method to exploit this rating information. MF techniques have gained popularity and become the standard recommender approaches due to their accuracy and scalability~\cite{koren09:MF}. They have probabilistic interpretation with Gaussian noise and are very flexible to add side data sources for recommender such as reviews content and social relations introduced in the following subsections. We adopt MF as a basic part of the proposed framework.

MF based RSs are mainly to find the latent user-specific matrix $U=[U_1,...,U_I] \in \mathbb{R}^{F \times I}$ and item-specific matrix $V=[V_1,...,V_J] \in \mathbb{R}^{F \times J}$, where $F$ is the number of latent factors, obtained by solving the following problem
\begin{equation}
\label{eq:rating}
\min_{U,V} \sum\nolimits_{R_{i,j} \neq 0} {(R_{i,j} - \hat R_{i,j})}^2 + \lambda (\norm {U}_F^2 + \norm {V}_F^2),
\end{equation}
where the predicted ratings
\begin{equation}
\label{eq:pred}
\hat R_{i,j} = \mu + b_i + b_j + U_i^{\textrm T} V_j,
\end{equation}
and regularization parameter $\lambda$ controls over-fitting. The rating mean is captured by $\mu$; $b_i$ and $b_j$ are rating biases of $u_i$ and of $v_j$. The $F$-dimensional feature vectors $U_i$ and $V_j$ represent preferences for user $i$ and characteristics for item $j$, respectively. The dot products $U_i^{\textrm T} V_j$ capture the interaction or match degree between users and items.

\subsection{Topic MF: Integrating Rating with Review}
Item reviews generated by users provide implicit feedback for recommender beyond explicit ratings~\cite{ganu09:review,TopicMF}. Reviews explain the ratings of users, thus help to understand the rating behavior of users, and alleviate the cold-item problem. On the one hand, item characteristics (i.e., factors) are latent in ratings, and can be found by MF introduced in Eq.(\ref{eq:rating}); on the other hand, item properties (i.e., topics) are hidden in reviews, and can be found by topic models like {\em latent Dirichlet allocation} (LDA)~\cite{LDA}. Together, these intuitions were sharpened into the HFT model~\cite{HFT}.

The HFT model combines ratings with reviews by minimizing the following problem
\begin{equation}
\label{eq:rev}
\sum_{R_{i,j} \neq 0} {(R_{i,j} - \hat R_{i,j})}^2 - \lambda \sum_{d=1}^J \sum_{n \in N_d} \log \theta_{z_{d,n}} \phi_{z_{d,n},w_{d,n}}
\end{equation}
where the LDA parameters $\theta$ and $\phi$ denote the topic and word distributions, respectively; $w_{d,n}$ and $z_{d,n}$ are the $n^{\mathrm{th}}$ word occurring in doc $d$ and the corresponding topic; and $\lambda$ controls the contribution from reviews content. Summation in the second term is over all documents and each word within.

The goals to achieve are both modeling ratings accurately and generating reviews likely. The trick of fusing ratings and reviews is the transformation
\begin{dmath}
\label{eq:tran}
\theta_{j,f} = \frac {\exp (\kappa V_{j,f})} {\sum_f \exp (\kappa V_{j,f})},
\end{dmath}
where the parameter $\kappa$ is introduced to control the `peakiness' of the transform and the summation is with respect to the $F$ latent topics/factors. The above function transforms the real-valued parameters $V_j \in \mathbb{R}^F$ associated with ratings to the probabilistic ones $\theta_j \in \Delta^F$ associated with reviews. The fusing trick works because if an item exhibits a certain property, it corresponds to some topic being commented by users. We adopt HFT as a component of the proposed framework.~\footnote{As the same with HFT, we aggregate all reviews of a particular item as a `doc'; so the item index $j$ is corresponding to doc index $j$.}

\subsection{Social MF: Integrating Rating with Relation}\label{paper:smf}
Social relations among users provide additional information for recommender~\cite{trustRS,tranMF}. On the one hand, social correlation theories~\cite{socialmedia} including homophily and social influence indicate that the rating behavior of users is correlated with their social factors hidden in the social network, besides their preference factors hidden in the rating matrix. On the other hand, the reputation of a user in the social network reveals her rating confidence, and a consideration from a global perspective can alleviate the rating noise to some extent. Together, these ideas were formulated in LOCABAL~\cite{LOCABAL}.

The LOCABAL model combines ratings with social relations to achieve the goals of modeling ratings accurately and capturing local social context by solving the problem
\begin{dmath}
\label{eq:locabal}
\min_{U,V,H} \sum\nolimits_{R_{i,j} \neq 0} W_{i,j} {(R_{i,j} - \hat R_{i,j})}^2 + \lambda \sum\nolimits_{T_{i,k} \neq 0} {(S_{i,k} - U_i^{\mathrm{T}} H U_k)}^2 + \lambda \Omega (\Theta),
\end{dmath}
where the rating weight $W_{i,j} = 1/(1+\log r_i)$ is computed from the PageRank score $r_i$ of user $i$ in the social network, representing the global perspective of social context; $S_{i,k}$ is the cosine similarity between rating vectors of user $i$ and $k$; $H \in \mathbb{R}^{F \times F}$ is the social correlation matrix, capturing the user preference correlation; $\lambda$ controls the contribution from social relations; and the regularization term is given by
\begin{equation}
\Omega (\Theta)=\norm {U}_F^2  + \norm {V}_F^2 + \norm{H}_F^2.
\end{equation}

\noindent
\textbf{eSMF.\quad} While LOCABAL succeeded in integrating ratings with social relations for recommender from local and global perspectives, it can be further improved by exploiting the graph structure of neighbors. Graph structure of neighbors captures social influence locality~\cite{localInfluence}, in other words, user behaviors are mainly influenced by direct friends in their ego networks. We employ the trust values used in SoRec~\cite{SoRec} to exploit this structure, and propose the {\em extended Social MF} (eSMF) model:
\begin{dmath}
\label{eq:esmf}
\min_{U,V,H} \sum\nolimits_{R_{i,j} \neq 0} W_{i,j} {(R_{i,j} - \hat R_{i,j})}^2 + \lambda \sum\nolimits_{T_{i,k} \neq 0} C_{i,k} {(S_{i,k} - U_i^{\mathrm{T}} H U_k)}^2 + \lambda \Omega (\Theta).
\end{dmath}
The trust values
\begin{dmath}
C_{ik} = \sqrt{d^-_{u_k} / (d^+_{u_i} + d^-_{u_k})},
\end{dmath}
where the outdegree $d^+_{u_i}$ represents the number of users whom $u_i$ trusts, while the indegree $d^-_{u_k}$ denotes the number of users who trust $u_k$.

\subsection{MR3: A Model of Rating, Review and Relation}{\label{paper:mr3}}
\begin{figure}
\centering
\includegraphics[height=4cm,width=3.1in]{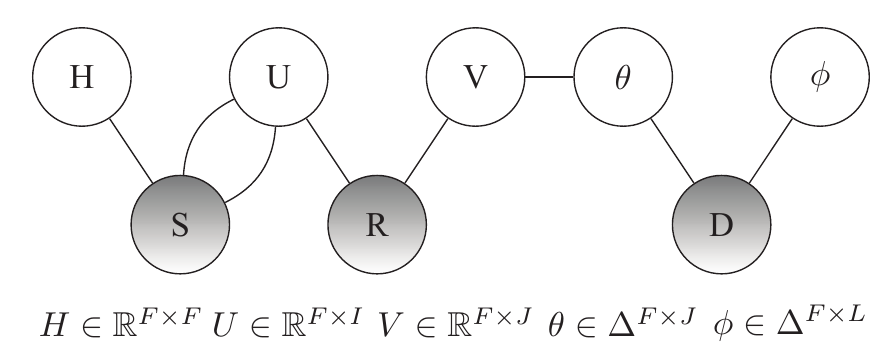}
\caption{ {\em Relationship among matrices of parameters and data.} Shaded nodes are data ($R$: rating matrix, $S$: social rating similarity, and $D$: doc-term matrix of reviews); Others are parameters ($U$: matrix of latent user factors, $V$: matrix of latent item factors, $H$: social correlation matrix, $\theta$: doc-topic distributions, and $\phi$: topic-word distributions). Parameters $V$ and $\theta$ are coupled by Eq.(\ref{eq:tran}). The double connections between $U$ and $S$ are indicated by the term $(S - U^{\textrm T}HU)$ in Eq.(\ref{eq:esmf}).}
\label{fig:depend}
\end{figure}

So far, we have described solutions to integrating ratings with reviews (see Eq.(\ref{eq:rev})) and to integrating ratings with social relations (see Eq.(\ref{eq:esmf})) based on MF respectively. By aligning latent factors and topics, we propose an effective framework {\mbox{MR3}} to jointly model ratings with social relations and reviews. MR3 connects Social MF and Topic MF by minimizing the following problem
\begin{multline}
\label{eq:mr3}
\mathcal{L}(\Theta,\Phi,z,\kappa) \triangleq \sum\nolimits_{R_{i,j} \neq 0} W_{i,j} {(R_{i,j} - \hat R_{i,j})}^2 \\
- \lambda_{\mathrm{rev}} \sum\nolimits_{d=1}^J \sum\nolimits_{n \in N_d} (\log \theta_{z_{d,n}} + \log \phi_{z_{d,n},w_{d,n}}) \\
+ \lambda_{\mathrm{rel}} \sum\nolimits_{T_{i,k} \neq 0} C_{i,k} {(S_{i,k} - U_i^{\textrm T} H U_k)}^2 + \lambda \Omega (\Theta) ,
\end{multline}
where parameters $\Theta = \{U,V,H\}$ are associated with ratings and social relations, parameters $\Phi =\{\theta,\phi\}$ associated with reviews text; and $\lambda_{\mathrm{rel}}$ and $\lambda_{\mathrm{rev}}$ are introduced to balance results from social relations and reviews, respectively.

Before we delve into the learning algorithm, a brief discussion on Eq.(\ref{eq:mr3}) is in order. On the right hand, the first term is the rating squared-error weighted by user reputation in the social network; the second term is the negative log likelihood of item reviews corpus; the third term is local social context factorization weighted by trust values among users; the last term is Frobenius norm penalty of parameters to control over-fitting. The connection between ratings and social relations is the shared user latent feature space $U$; ratings and reviews are linked through the transformation involving $V$ and $\theta$ in Eq.(\ref{eq:tran}). The dependencies among these data and parameter matrices are depicted in Figure \ref{fig:depend}.

\noindent
\textbf{Learning.\quad} Our objective is to search
\begin{dmath}
\label{eq:mr3f}
\argmin_{\Theta,\Phi,z,\kappa} \mathcal{L}(\Theta,\Phi,z,\kappa).
\end{dmath}
Observe that parameters $\Theta$ and $\Phi$ are coupled (see above paragraph, Eq.(\ref{eq:tran}), or Figure \ref{fig:depend}). The former can be found by gradient descent and the latter by Gibbs sampling; so, we design a procedure alternating between following two steps:
\begin{subequations}
\label{eq:2step}
\begin{equation}
\label{eq:step1}
 \mbox{update } \Theta^{\mathrm{new}}, \Phi^{\mathrm{new}},\kappa^{\mathrm{new}} = \argmin_{\Theta,\Phi,\kappa} \mathcal{L}(\Theta,\Phi,\kappa,z^{\mathrm{old}});
 \end{equation}
\begin{equation}
\label{eq:step2}
 \mbox{sample }\; z_{d,n}^{\mathrm{new}} \mbox{ with probability } p(z_{d,n}^{\mathrm{new}} = f) = \phi_{f,w_{d,n}}^{\mathrm{new}}.
\end{equation}
\end{subequations}

For the first step Eq.(\ref{eq:step1}), topic assignments $z_{d,n}$ for each word in reviews corpus are fixed; then we update the terms {$\Theta,\Phi$, and $\kappa$} by gradient descent (GD). Recall that $\theta$ and $V$ depend on each other; we fit only $V$ and then determine $\theta$ by Eq.(\ref{eq:tran}). This is the same as that in the standard gradient-based MF for recommender except that we have to compute more gradients, which will be given later separately.

For the second step Eq.(\ref{eq:step2}), parameters associated with reviews corpus $\theta$ and $\phi$ are fixed; then we sample topic assignments $z_{d,n}$ by iterating through all docs $d$ and each word within, setting $z_{d,n} = f$ with probability proportion to $\theta_{d,f} \phi_{f,w_{d,n}}$. This is similar to updating $z$ via LDA except that topic proportions $\theta$ are not sampled from a Dirichlet prior, but instead are determined in the first step.

Finally, the two steps are repeated until a local optimum is reached. In practice, we sample topic assignments every 5 GD iterations/epoches and this is called a pass; usually it is enough to run 50 passes to find a local minima.

\noindent
\textbf{Gradients.\quad} We now give gradients used in Eq.(\ref{eq:step1}). (Gradients of biases are omitted; rating mean is not fitted because ratings are centered.) More notations are required here~\cite{Gibbs}. For each item $j$ (i.e. doc $j$): 1) $M_j$ is an $F$-dimensional count vector, in which each component is the number of times each topic occurs for it; 2) $m_j$ is the number of words in it; and 3) $z_j = \sum\nolimits_f \exp{(\kappa V_{jf})}$ is a normalizer. For each word $w$: 1) $M_w$ is an $F$-dimensional count vector, in which each component is the number of times it has been assigned to each topic; 2) $m_f$ is the number of times topic $f$ occurs; and 3)$z_f = \sum\nolimits_w \exp{(\psi_{fw})}$ is a normalizer. Note that $\phi_f$ is a stochastic vector, so we optimize the corresponding unnormalized vector $\psi_f$ and then get $\phi_{fw} = \exp{(\psi_{fw})}/z_f$.
\begin{multline}
\label{eq:grad-U}
 \frac 1 2 \frac {\partial\mathcal{L}} {\partial U_i} = \sum\nolimits_{j:R_{i,j} \neq 0} W_{i,j}(\hat R_{i,j} - R_{i,j})V_j + \lambda U_i \\
 + \lambda_{\mathrm{rel}} \sum\nolimits_{k:T_{k,i} \neq 0} C_{i,k}(U_k^{\mathrm T}HU_i - S_{i,k})H^{\mathrm T}U_k \\
 + \lambda_{\mathrm{rel}} \sum\nolimits_{k:T_{i,k} \neq 0} C_{k,i}(U_i^{\mathrm T}HU_k - S_{i,k})HU_k .
\end{multline}
\begin{dmath}
\label{eq:grad-V}
\frac {\partial\mathcal{L}} {\partial V_j} = 2 \sum\nolimits_{i:R_{i,j} \neq 0} W_{i,j}(\hat R_{i,j} - R_{i,j})U_i \\
 - \lambda_{\mathrm{rev}} \kappa \Big(M_j - \frac {m_j} {z_j} \exp{(\kappa V_j})\Big) + 2 \lambda V_j.
\end{dmath}
\begin{dmath}
\label{eq:grad-H}
\frac 1 2 \frac {\partial\mathcal{L}} {\partial H} = \lambda_{\mathrm{rel}} \sum_{T_{i,k} \neq 0} C_{i,k}(U_i^{\mathrm T}HU_k - S_{i,k})U_i U_{k}^{\mathrm{T}} + \lambda H .
\end{dmath}
\begin{equation}
\label{eq:grad-phi}
\frac {\partial\mathcal{L}} {\partial \psi_{fw}} = - \lambda_{\mathrm{rev}} \Big(M_{fw} - \frac {m_f} {z_f} \exp{(\psi_{fw}})\Big).
\end{equation}
\begin{equation}
\label{eq:grad-kappa}
\frac {\partial\mathcal{L}} {\partial \kappa} = - \lambda_{\mathrm{rev}} \sum\nolimits_{j,f} V_{jf} \Big(M_{jf} - \frac{m_j}{z_j} \exp{(\kappa V_{jf}})\Big).
\end{equation}

\section{Experiments}\label{paper:Exp}
In this section, we first evaluate our proposed eSMF component to show the benefit of exploiting the graph structure of neighbors. Then we demonstrate the effectiveness of our proposed MR3 model compared with the individual components. Finally we analyze the contribution of each component of data source to the proposed model, followed by sensitivity of MR3 to hyperparameters.
\subsection{Datasets and Metric}
We evaluate our models on two datasets: Epinions and Ciao.\footnote{ \url{http://www.public.asu.edu/~jtang20/} } They are both knowledge sharing and review sites, in which users can rate items, connect to others, and give reviews on products. We remove stop words\footnote{\url{http://www.ranks.nl/stopwords}} and then select top $L$ = 8000 frequent words as vocabulary; we remove users and items that occur only once or twice. The items indexed in the rating matrix are aligned to documents in the doc-term matrix, that is, we aggregate all reviews of a particular item as a `doc'. Statistics of datasets are given in Table~\ref{table:data}. We see that the rating matrices of both datasets are very sparse, and the average length of documents is short on Epinions.
\begin{table}[h]
\centering
\begin{tabular}{l c c}
 \Xhline{2\arrayrulewidth}
    Statistics          & Epinions  & Ciao \\
 \hline \hline
 \# of Users            & 49,454    & 7,340\\
 \# of Items            & 74,154    & 22,472\\
 \# of Ratings/Reviews  & 790,940   & 183,974\\
 \# of Social Relations & 434,680   & 112,942\\
 \# of Words            & 2,246,837 & 28,874,000\\
 Rating Density         & 0.00022   & 0.0011\\
 Social Density         & 0.00018   & 0.0021\\
 Ave. Words Per Item    & 30.3      & 1284.9\\
 \Xhline{2\arrayrulewidth}
\end{tabular}
\caption{Statistics of the Two Datasets}
\label{table:data}
\end{table}

We randomly select $x$\% as the training set and report the prediction performance on the remaining 1 - $x$\% testing set. The metric {\em root-mean-square error} (RMSE) for rating prediction task is defined as
\begin{equation}
RMSE_{\mathcal{T}} = \sqrt { {\sum\nolimits_{(u_i,v_j) \in \mathcal{T}} (R_{i,j} - \hat R_{i,j})^2} \Big{/} {|\mathcal{T}|} }
\end{equation}
where $\mathcal{T}$ and $|\mathcal{T}|$ is the test set and its cardinality. A smaller RMSE means a better prediction performance.

\subsection{Comparing Social MF Methods}
\begin{figure}[h]
\centering
\subfigure{ \includegraphics[height=3.2cm,width=1.6in]{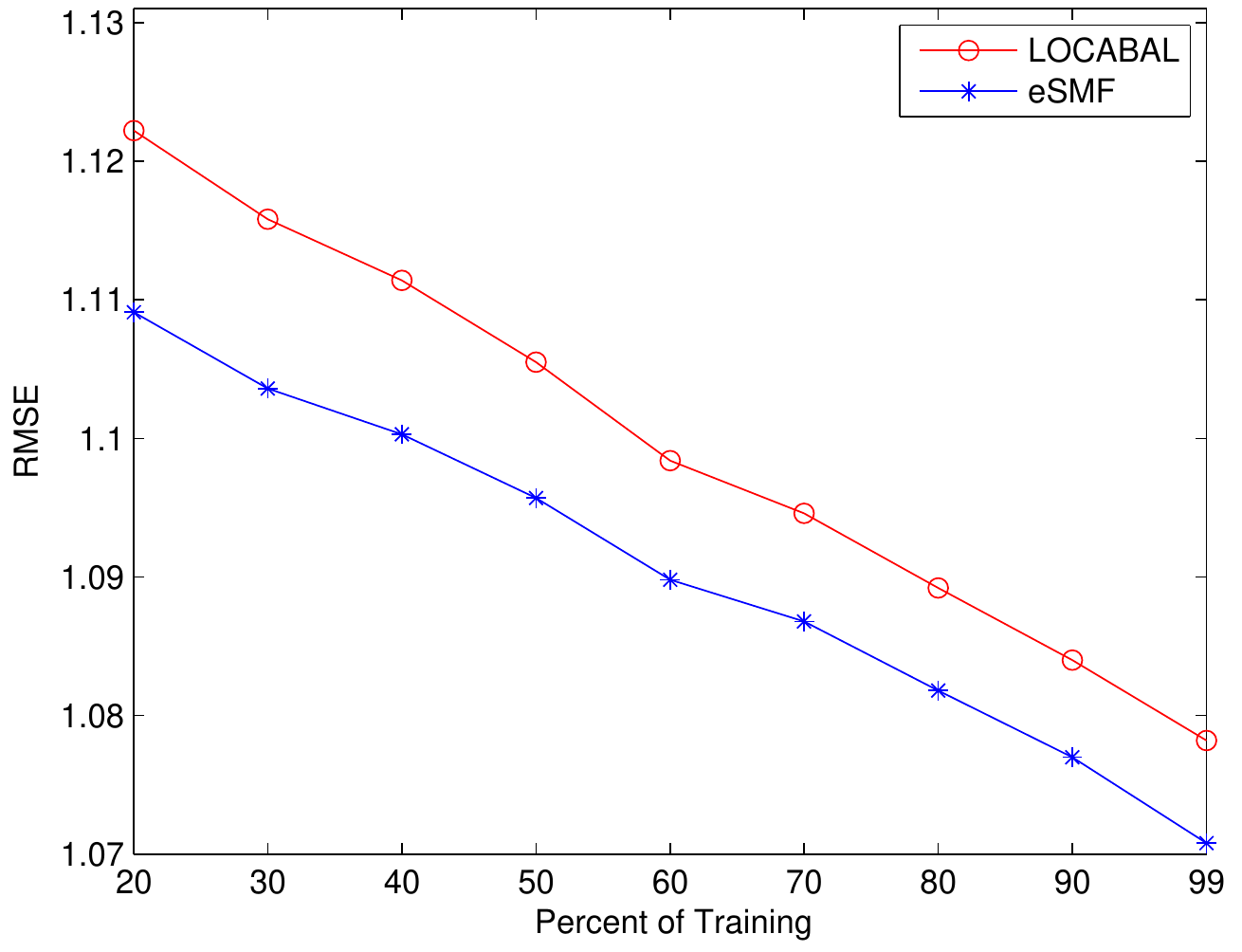} }
\subfigure{ \includegraphics[height=3.2cm,width=1.6in]{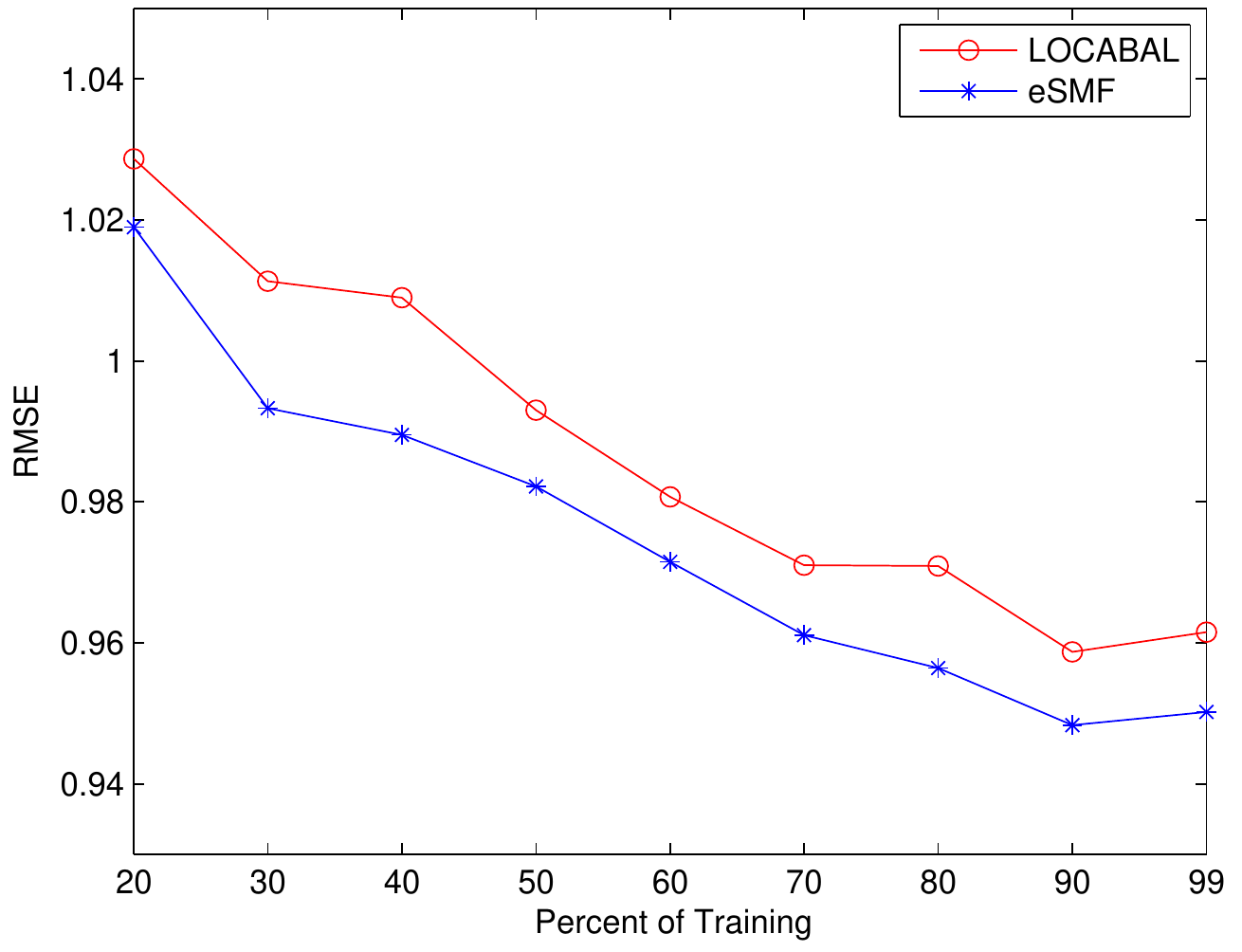} }
\caption{ {\em Comparisons of eSMF and LOCABAL on two datasets.} Left: Epinions; Right: Ciao.}
\label{fig:eSMF}
\end{figure}

We first compare the eSMF method introduced in Subsec~\ref{paper:smf} with LOCABAL~\cite{LOCABAL}, a recent Social MF method. The motivation for the comparison is two-fold: 1) to demonstrate that exploiting ratings and social relations more tightly can further improve the performance of social RSs; 2) to form a nice component of the framework {\mbox{MR3}}, which we will evaluate in the following subsection.

We use grid search to determine hyperparameters and report the best RMSE on the testing set over 50 passes (the same routing for comparing MR3 below). For both eSMF and LOCABAL, the number of latent factors $F = 10$, norm penalty $\lambda = 0.5$, learning rate = 0.0007, momentum = 0.8, and $\lambda_{\mathrm{rel}} = 0.1$. Parameters $\Theta= \{U,V,H\}$ are randomly initialized from $\mathcal{N}(0,0.01)$. The results are given in Figure~\ref{fig:eSMF}, with varying percentage of the training set = \{20, 30, 40, 50, 60, 70, 80, 90, 99\} and we have the following observation:
\begin{itemize}
\item {Exploiting ratings and social relations tightly can further improve recommender performance in terms of RMSE on both datasets. For example, eSMF obtains 1.18\%, 0.89\%, and 0.72\% relative improvement on Epinions with 20\%, 50\%, and 70\% as the training set respectively.}
\end{itemize}

\subsection{Comparing Different Recommender Systems}
\begin{table*}[t]
\centering
\begin{tabular}{ccccccc|ccc}
\hline \hline
\multicolumn{1}{c}{\multirow{2}{*}{Datasets}} & \multirow{2}{*}{Training} & \multicolumn{5}{c|}{Methods}   & \multicolumn{3}{c}{Improvement of MR3 vs. } \\ \cline{3-7} \cline{8-10}
\multicolumn{1}{c}{}                          &        & Mean   & PMF    & HFT    & LOCABAL & MR3    & PMF          & HFT          & LOCABAL      \\
\hline \hline
\multirow{4}{*}{Epinions} & 20\% & 1.2265 & 1.2001 & 1.1857 & 1.1222  & 1.1051 & 8.60\% & 7.29\% & 1.55\%\\
                          & 50\% & 1.2239 & 1.1604 & 1.1323 & 1.1055  & 1.0809 & 7.35\% & 4.76\% & 2.28\%\\
                          & 80\% & 1.2225 & 1.1502 & 1.0960 & 1.0892  & 1.0648 & 8.02\% & 2.93\% & 2.29\%\\
                          & 90\% & 1.2187 & 1.1484 & 1.0867 & 1.0840  & 1.0634 & 7.99\% & 2.19\% & 1.94\%\\
\hline \hline
\multirow{4}{*}{Ciao}   & 20\% & 1.1095 & 1.0877 & 1.0439 & 1.0287  & 1.0142 & 7.25\% & 2.93\% & 1.43\%\\
                        & 50\% & 1.0964 & 1.0536 & 1.0379 & 0.9930  & 0.9740 & 8.17\% & 6.56\% & 1.95\%\\
                        & 80\% & 1.0899 & 1.0418 & 0.9958 & 0.9709  & 0.9521 & 9.42\% & 4.59\% & 1.97\%\\
                        & 90\% & 1.0841 & 1.0391 & 0.9644 & 0.9587  & 0.9451 & 9.95\% & 2.04\% & 1.44\%\\
\hline \hline
{Average} & & & & & & & 8.34\% & 4.16\% & 1.86\% \\
\hline \hline
\end{tabular}
\captionsetup{justification=centering}
\caption{RMSE Comparisons of Different Methods ($F = 10$)}
\label{table:mr3}
\end{table*}

In this subsection, we compare the proposed framework MR3 introduced in Subsec~\ref{paper:mr3} with the following baselines:

\textbf{Mean.\quad} This method predicts the rating always using the average, i.e. $\mu$ in Eq.(\ref{eq:pred}), across all training ratings. This is the best constant predictor in terms of RMSE.

\textbf{PMF.\quad} This method performs matrix factorization on rating matrix as shown in Eq.(\ref{eq:rating})~\cite{PMF}. It only uses the rating source.

\textbf{HFT.\quad} This method combines latent factors in ratings with hidden topics in reviews as shown in Eq.(\ref{eq:rev})~\cite{HFT}. It only uses ratings and reviews.

\textbf{LOCABAL.\quad} This method is based on matrix factorization and exploits local and global social context as shown in Eq.(\ref{eq:locabal})~\cite{LOCABAL}. It only uses ratings and relations.

We use the source code PMF\footnote{\url{http://www.cs.toronto.edu/~rsalakhu/}} and HFT\footnote{\url{http://cseweb.ucsd.edu/~jmcauley/}}. For all methods, we set the number of latent factors $F = 10$, norm penalty $\lambda = 0.5$, learning rate = 0.0007, momentum = 0.8. For HFT, $\lambda_{\mathrm{rev}} = 0.1$; for MR3, $\lambda_{\mathrm{rel}} = 0.001$ and $\lambda_{\mathrm{rev}} = 0.05$. More details about the sensitivity to parameters of MR3 will be discussed later. The results of the comparison are summarized in Table~\ref{table:mr3} and we have the following observations.
\begin{itemize}
\item {Exploiting social relations and reviews beyond ratings can significantly improve recommender performance in terms of RMSE on both datasets. For example, HFT and LOCABAL obtain 4.95\% and 5.60\% relative improvement compared with PMF on Epinions with 80\% as the training set respectively.}
\item {Our proposed framework MR3 always achieves the best result. Compared with HFT and LOCABAL, MR3 averagely gains 0.0466 and 0.0217 absolute RMSE improvement on Epinions and 0.0392 and 0.0165 on Ciao respectively. The main reason is that MR3 jointly models all three types of information. The contribution from each data source to MR3 is discussed in the following subsection.}
\end{itemize}

\subsection{Impact of Social Relations and Reviews}
\begin{figure}[h]
\centering
\subfigure{ \includegraphics[height=3.4cm,width=1.6in]{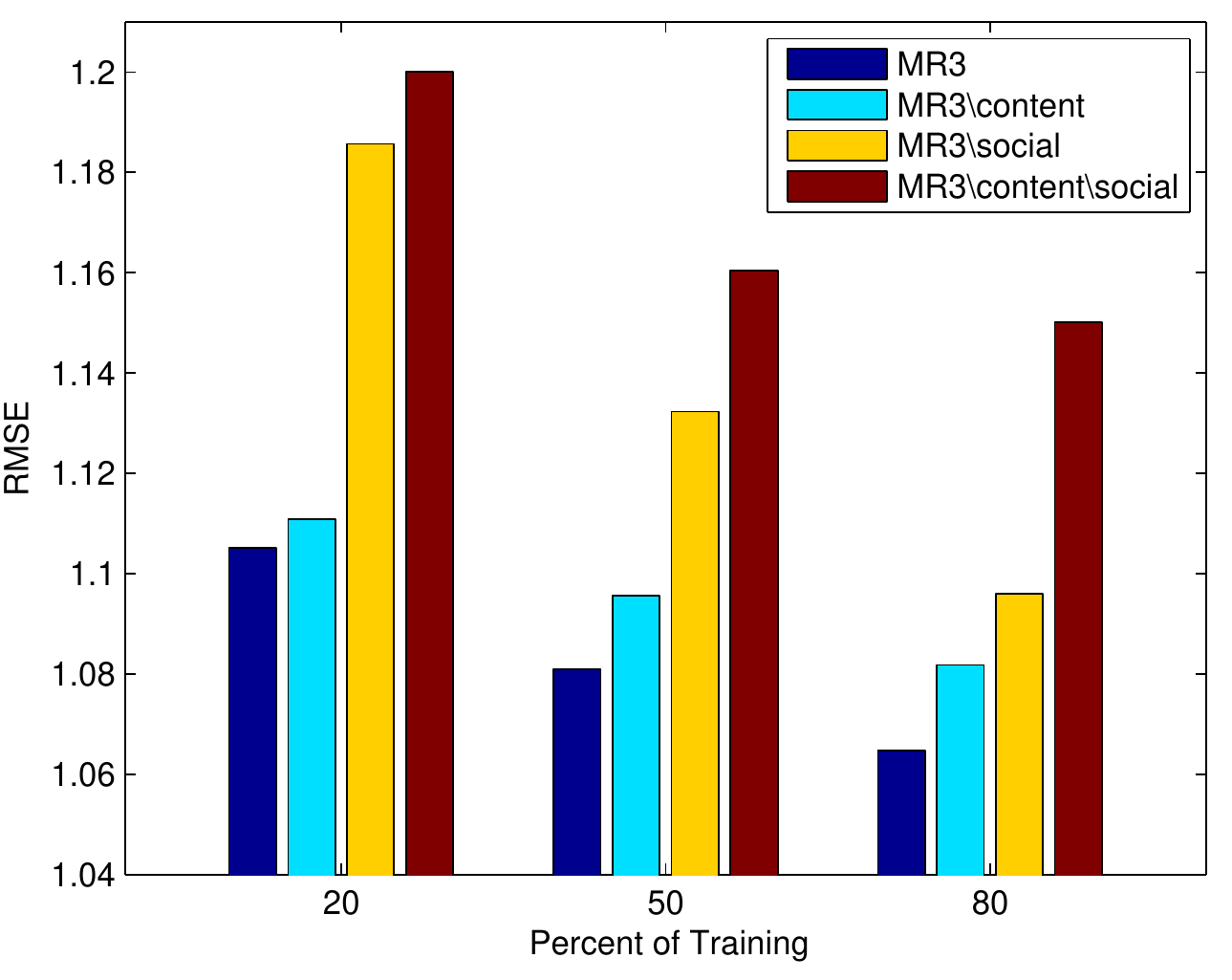} }
\subfigure{ \includegraphics[height=3.4cm,width=1.6in]{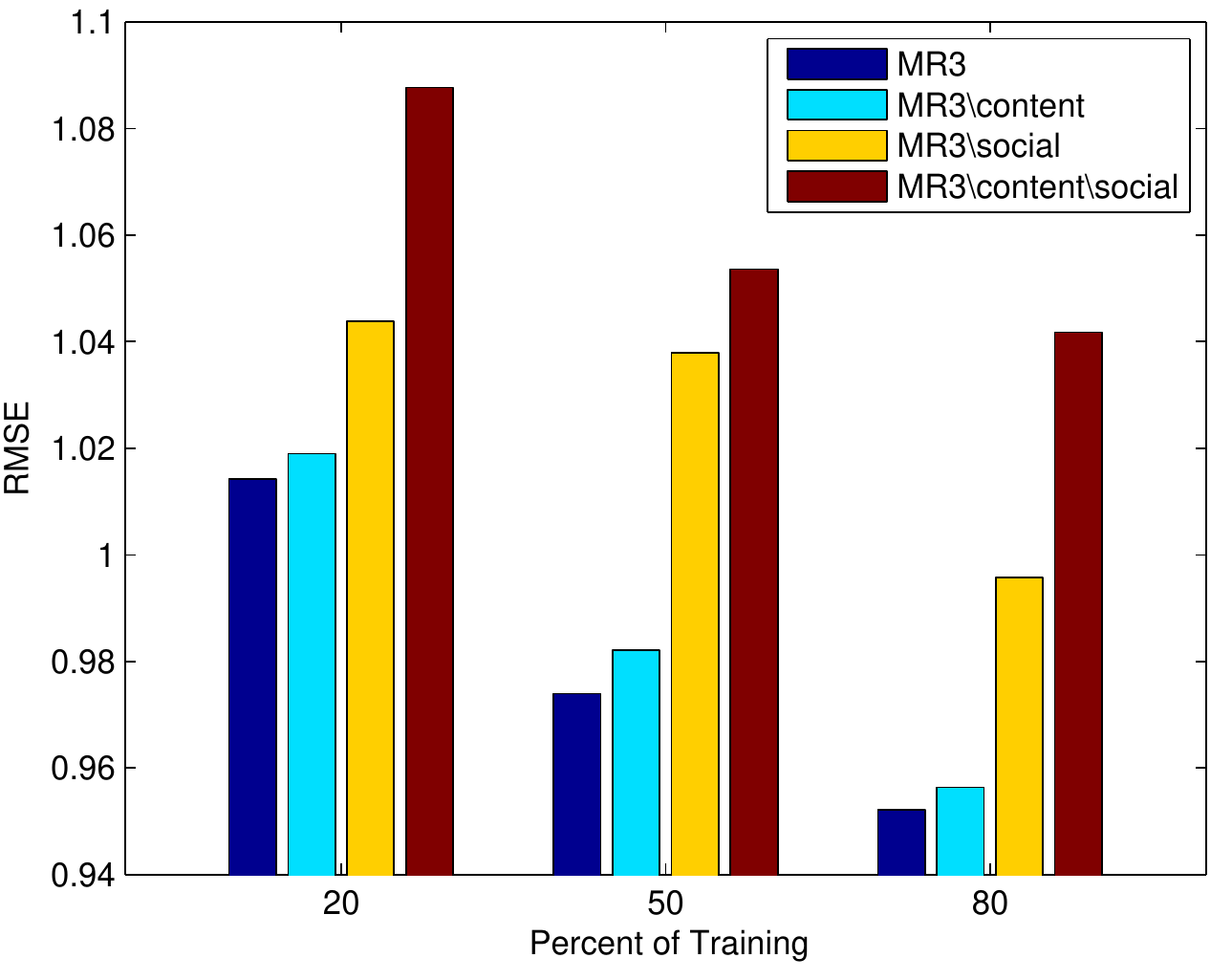} }
\caption{ {\em Predictive performance of MR3 compared with its three components.} Left: Epinions; Right: Ciao.}
\label{fig:component}
\end{figure}

We have shown the effectiveness of integrating ratings with social relations and reviews in our proposed framework MR3. We now investigate the contribution of each data source to the MR3 by eliminating the impact of social relations and reviews from it in turn:

\textbf{MR3$\backslash$content:\quad} Eliminating the impact of reviews by setting $\lambda_{\mathrm{rev}} = 0$ in Eq.(\ref{eq:mr3}), which is equivalent to eSMF as shown in Eq.(\ref{eq:esmf}).

\textbf{MR3$\backslash$social:\quad} Eliminating the impact of social relations by setting $\lambda_{\mathrm{rel}} = 0$ in Eq.(\ref{eq:mr3}), which is equivalent to HFT as shown in Eq.(\ref{eq:rev}).

\textbf{MR3$\backslash$content$\backslash$social:\quad} Eliminating the impact of both reviews and social relations by setting $\lambda_{\mathrm{rev}} = 0$ and $\lambda_{\mathrm{rel}} = 0$ in Eq.(\ref{eq:mr3}), which is equivalent to PMF as shown in Eq.(\ref{eq:rating}).

The predictive results of MR3 and its three components are shown in Figure~\ref{fig:component}. The performance degrades when either social relations or reviews are eliminated. In detail, {\em \mbox{MR3$\backslash$content}}, {\em \mbox {MR3$\backslash$social}}, and {\em \mbox{MR3$\backslash$content$\backslash$social}} averagely reduce 1.19\%, 4.29\%, and 7.99\% relative RMSE performance on Epinions respectively, suggesting that both reviews and social relations contain essential information for recommender.

\subsection{Sensitivity to Parameters: $F$, $\lambda_{\mathrm{rel}}$ and $\lambda_{\mathrm{rev}}$}
The framework MR3 has three important hyperparameter: 1) the number of latent factors $F$; 2) the $\lambda_{\mathrm{rev}}$ that controls the contribution from reviews; and 3) the $\lambda_{\mathrm{rel}}$ that controls the contribution from social relations. We investigate the sensitivity of MR3 to these parameters by varying one of them while fixing the other two.

\begin{figure}[!h]
\centering
\subfigure{\includegraphics[height=3.4cm,width=1.66in]{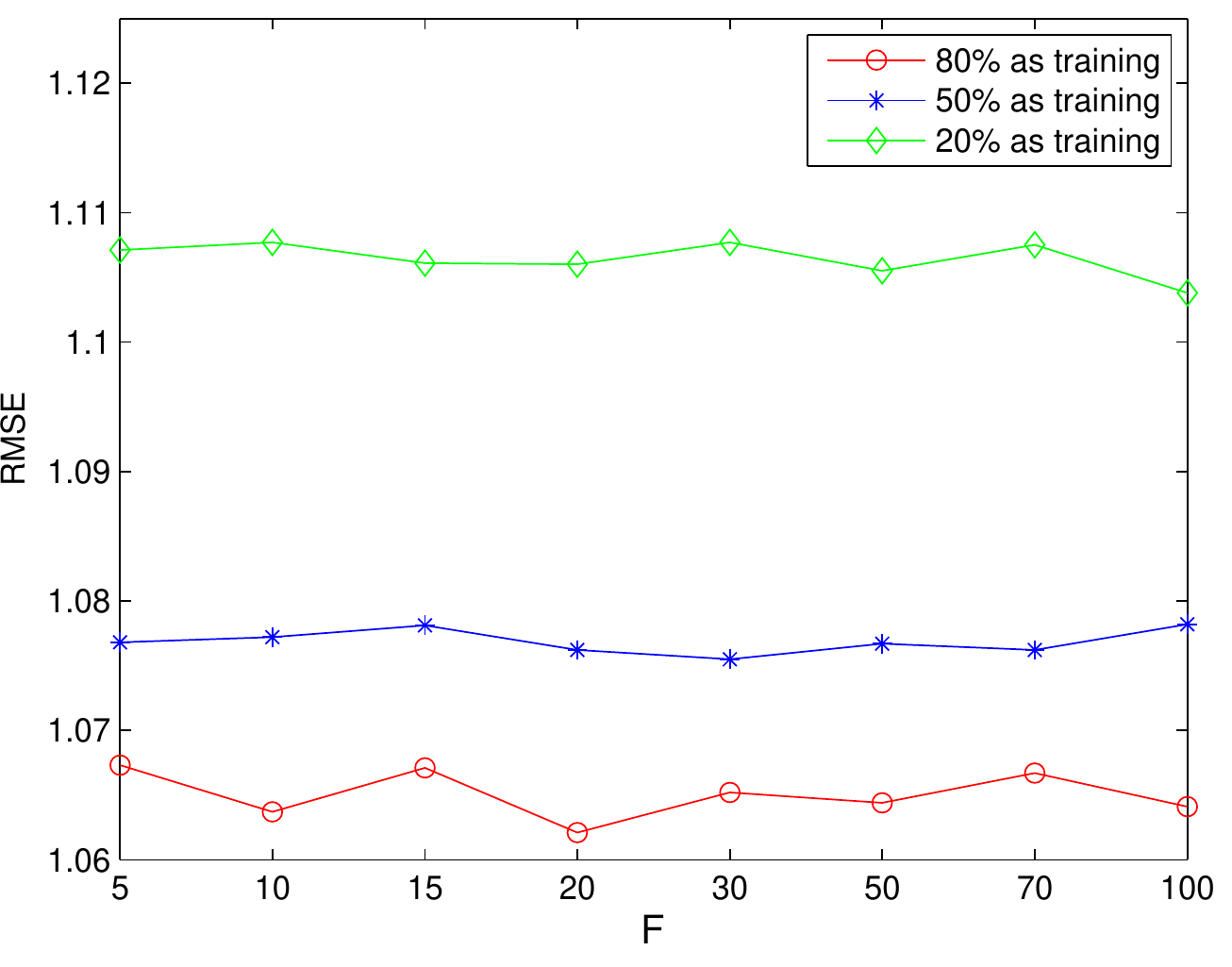}}
\subfigure{\includegraphics[height=3.4cm,width=1.66in]{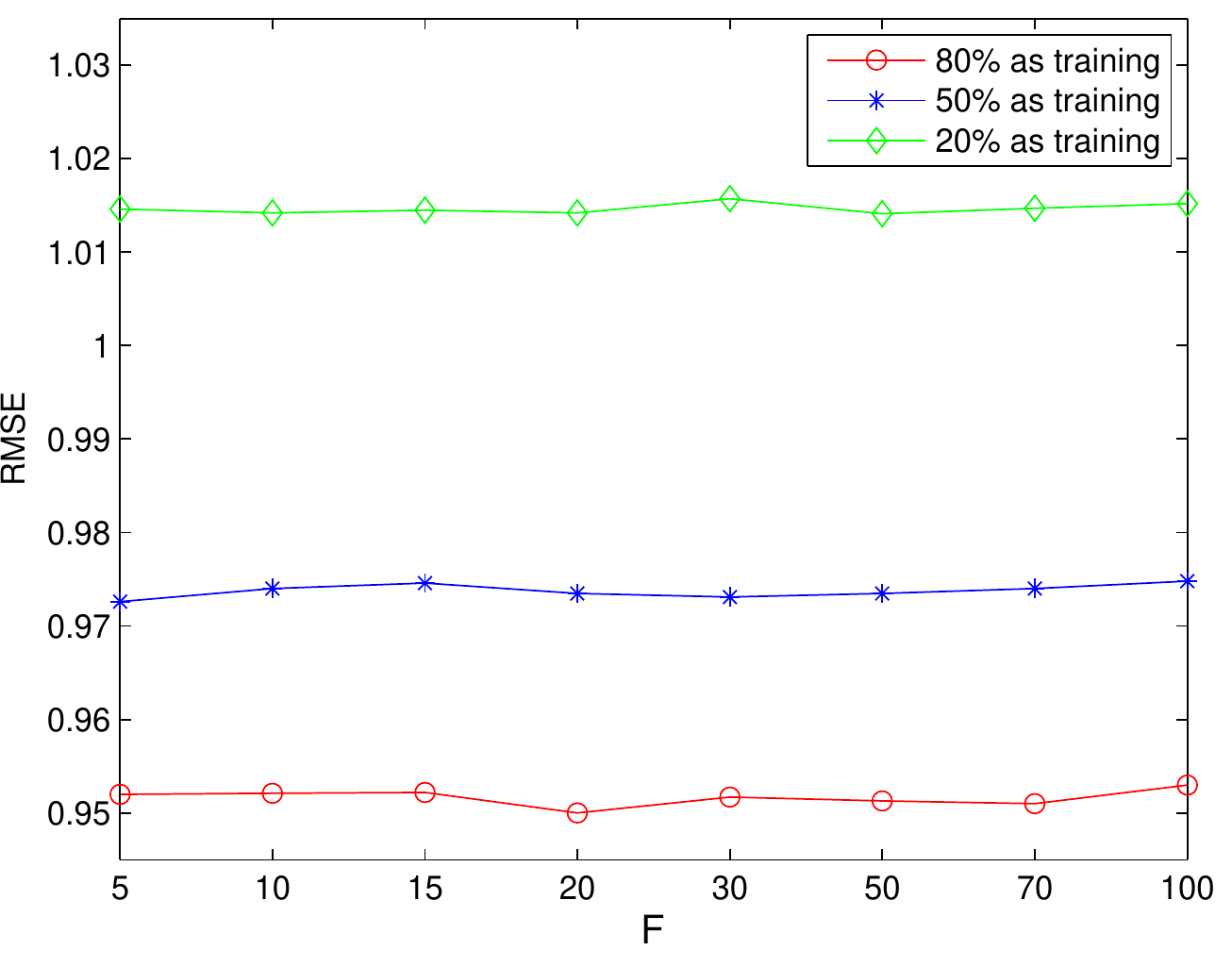}}
\caption{ {\em Predictive performance of MR3 by varying the number of latent factors $F$.} Fixing $\lambda_{\mathrm{rel}} = 0.001$ and $\lambda_{\mathrm{rev}} = 0.05$. Left: Epinions; Right: Ciao. }
\label{fig:mr3k}
\end{figure}

First, we fix $\lambda_{\mathrm{rel}} = 0.001$ and $\lambda_{\mathrm{rev}} = 0.05$, and vary the number of latent factors $F = \{5, 10, 15, 20, 30, 50, 70, 100\}$ with 20\%, 50\%, 80\% as the training set respectively. As shown in Figure~\ref{fig:mr3k}, MR3 is relatively stable and not sensitive to $F$, so we choose the reasonable value 10 as default.

\begin{figure}[h]
\centering
\includegraphics[height=5cm,width=2.5in]{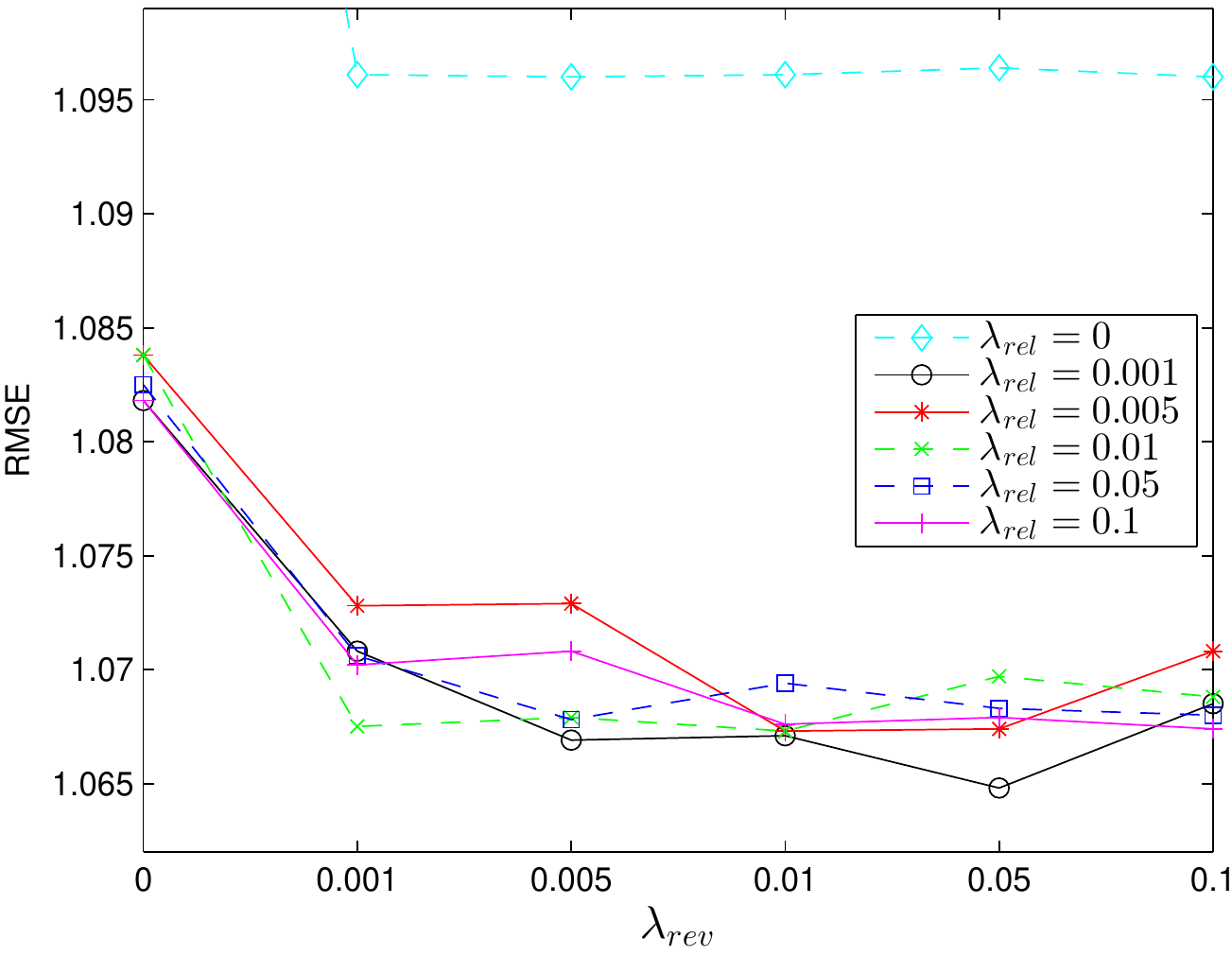}
\caption{ {\em Predictive performance of MR3 by varying $\lambda_{\mathrm{rel}}$ and $\lambda_{\mathrm{rev}}$.} Both vary in \{0, 0.001, 0.005, 0.01, 0.05, 0.1\}. RMSE is 1.1502 when both are zero. Fixing $F$ = 10. Percent of training set = 80. Dataset: Epinions. }
\label{fig:mr3relu}
\end{figure}

Next, we fix $F = 10$ and study how the reviews associated hyperparameter $\lambda_{\mathrm{rev}}$ and the social relations associated one $\lambda_{\mathrm{rel}}$ affect the whole performance of MR3.  As shown in Figure~\ref{fig:mr3relu}, we have some observations: 1) the prediction performance degrades when either $\lambda_{\mathrm{rel}} = 0$ or $\lambda_{\mathrm{rev}} = 0$; (RMSE is 1.1502 when both are zero.) 2) MR3 is relatively stable and not sensitive to $\lambda_{\mathrm{rel}} $ and $\lambda_{\mathrm{rev}}$ when they are small (e.g., from 0.0001 to 0.1), so we choose the reasonable values 0.001 and 0.05 for them respectively.

\section{Conclusion and Future Work}
Heterogenous recommending information sources beyond explicit ratings like social relations and item reviews present both opportunities and challenges for conventional recommender systems. We investigate how to fuse these three kinds of information tightly and effectively for recommendation. A unified framework \mbox{MR3} by aligning latent factors and topics is proposed to perform social matrix factorization and topic matrix factorization simultaneously for effective rating prediction. Empirical results on real-world datasets show that our proposed model leads to improved predictive performance. Further experiments are designed to see the impact of each of the data sources.

The proposed model has some limitations which provide interesting directions for future work. Typically, the number of hidden topics in reviews is less than that of latent factors in ratings; therefore the assumption that these two are equal in the current model is inappropriate~\cite{JMARS}. As inclination of users, popularity of items, and structure of the social network constantly change, integrating temporal dynamics into MR3 is worth channeling. Integrating Implicit feedback should further improve the performance~\cite{SVDPP}. Recently, deep neural networks (i.e., deep learning) have been used to learn better representation of both items' characteristics and content for recommendation~\cite{dl4rs}, so the issue of integrating them into MR3 framework as a nicer component is also interesting.

\newpage
\section*{Acknowledgement}
We thank Jiliang Tang for providing datasets. The work was supported by NSFC (61472183, 61333014) and 863 program(2015AA015406).

\bibliographystyle{named}
\bibliography{ijcai15}

\begin{thebibliography}{}

\bibitem[\protect\citeauthoryear{Bao \bgroup \em et al.\egroup
  }{2014}]{TopicMF}
Yang Bao, Hui Fang, and Jie Zhang.
\newblock Topicmf: Simultaneously exploiting ratings and reviews for
  recommendation.
\newblock In {\em AAAI}, 2014.

\bibitem[\protect\citeauthoryear{Bedi \bgroup \em et al.\egroup
  }{2007}]{trustRS}
Punam Bedi, Harmeet Kaur, and Sudeep Marwaha.
\newblock Trust based recommender system for semantic web.
\newblock In {\em IJCAI}, 2007.

\bibitem[\protect\citeauthoryear{Billsus and Pazzani}{1998}]{CFSVD98}
Daniel Billsus and Michael~J Pazzani.
\newblock Learning collaborative information filters.
\newblock In {\em ICML}, 1998.

\bibitem[\protect\citeauthoryear{Blei \bgroup \em et al.\egroup }{2003}]{LDA}
David~M Blei, Andrew~Y Ng, and Michael~I Jordan.
\newblock Latent dirichlet allocation.
\newblock {\em JMLR}, 2003.

\bibitem[\protect\citeauthoryear{Chen \bgroup \em et al.\egroup
  }{2014}]{CCTRSoRec}
Chaochao Chen, Xiaolin Zheng, Yan Wang, Fuxing Hong, and Zhen Lin.
\newblock Context-aware collaborative topic regression with social matrix
  factorization for recommender systems.
\newblock In {\em AAAI}, 2014.

\bibitem[\protect\citeauthoryear{Diao \bgroup \em et al.\egroup }{2014}]{JMARS}
Qiming Diao, Minghui Qiu, Chao-Yuan Wu, Alexander~J Smola, Jing Jiang, and
  Chong Wang.
\newblock Jointly modeling aspects, ratings and sentiments for movie
  recommendation (jmars).
\newblock In {\em KDD}, 2014.

\bibitem[\protect\citeauthoryear{Ganu \bgroup \em et al.\egroup
  }{2009}]{ganu09:review}
Gayatree Ganu, Noemie Elhadad, and Am{\'e}lie Marian.
\newblock Beyond the stars: Improving rating predictions using review text
  content.
\newblock In {\em WebDB}, 2009.

\bibitem[\protect\citeauthoryear{Goldberg \bgroup \em et al.\egroup
  }{1992}]{CF92}
David Goldberg, David Nichols, Brian~M Oki, and Douglas Terry.
\newblock Using collaborative filtering to weave an information tapestry.
\newblock {\em CACM}, 1992.

\bibitem[\protect\citeauthoryear{Griffiths and Steyvers}{2004}]{Gibbs}
Thomas~L Griffiths and Mark Steyvers.
\newblock Finding scientific topics.
\newblock {\em PNAS}, 2004.

\bibitem[\protect\citeauthoryear{Guo \bgroup \em et al.\egroup
  }{2015}]{TrustSVD}
Guibing Guo, Jie Zhang, and Neil Yorke-Smith.
\newblock Trustsvd: Collaborative filtering with both the explicit and implicit
  influence of user trust and of item ratings.
\newblock In {\em AAAI Conference on Artificial Intelligence}, 2015.

\bibitem[\protect\citeauthoryear{Jakob \bgroup \em et al.\egroup
  }{2009}]{jakob09review}
Niklas Jakob, Stefan~Hagen Weber, Mark~Christoph M{\"u}ller, and Iryna
  Gurevych.
\newblock Beyond the stars: exploiting free-text user reviews to improve the
  accuracy of movie recommendations.
\newblock In {\em CIKM workshop}, 2009.

\bibitem[\protect\citeauthoryear{Jamali and Ester}{2011}]{tranMF}
Mohsen Jamali and Martin Ester.
\newblock A transitivity aware matrix factorization model for recommendation in
  social networks.
\newblock In {\em IJCAI}, 2011.

\bibitem[\protect\citeauthoryear{Koren \bgroup \em et al.\egroup
  }{2009}]{koren09:MF}
Yehuda Koren, Robert Bell, and Chris Volinsky.
\newblock Matrix factorization techniques for recommender systems.
\newblock {\em Computer, IEEE}, 42(8):30--37, 2009.

\bibitem[\protect\citeauthoryear{Koren}{2008}]{SVDPP}
Yehuda Koren.
\newblock Factorization meets the neighborhood: a multifaceted collaborative
  filtering model.
\newblock In {\em KDD}, 2008.

\bibitem[\protect\citeauthoryear{Linden \bgroup \em et al.\egroup
  }{2003}]{linden03:amazon}
Greg Linden, Brent Smith, and Jeremy York.
\newblock Amazon.com recommendations: Item-to-item collaborative filtering.
\newblock {\em Internet Computing, IEEE}, 7(1):76--80, 2003.

\bibitem[\protect\citeauthoryear{Ling \bgroup \em et al.\egroup }{2014}]{RMR}
Guang Ling, Michael~R Lyu, and Irwin King.
\newblock Ratings meet reviews, a combined approach to recommend.
\newblock In {\em RecSys}, 2014.

\bibitem[\protect\citeauthoryear{Ma \bgroup \em et al.\egroup }{2008}]{SoRec}
Hao Ma, Haixuan Yang, Michael~R Lyu, and Irwin King.
\newblock Sorec: Social recommendation using probabilistic matrix
  factorization.
\newblock In {\em CIKM}, 2008.

\bibitem[\protect\citeauthoryear{Ma \bgroup \em et al.\egroup }{2011}]{SoReg}
Hao Ma, Dengyong Zhou, Chao Liu, Michael~R Lyu, and Irwin King.
\newblock Recommender systems with social regularization.
\newblock In {\em WSDM}, 2011.

\bibitem[\protect\citeauthoryear{Marsden and Friedkin}{1993}]{influence}
Peter~V Marsden and Noah~E Friedkin.
\newblock Network studies of social influence.
\newblock {\em Sociological Methods \& Research}, 22(1):127--151, 1993.

\bibitem[\protect\citeauthoryear{McAuley and Leskovec}{2013}]{HFT}
Julian McAuley and Jure Leskovec.
\newblock Hidden factors and hidden topics: understanding rating dimensions
  with review text.
\newblock In {\em RecSys}, 2013.

\bibitem[\protect\citeauthoryear{McPherson \bgroup \em et al.\egroup
  }{2001}]{homophily}
Miller McPherson, Lynn Smith-Lovin, and James~M Cook.
\newblock Birds of a feather: Homophily in social networks.
\newblock {\em Annual review of sociology}, 2001.

\bibitem[\protect\citeauthoryear{Mnih and Salakhutdinov}{2007}]{PMF}
Andriy Mnih and Ruslan Salakhutdinov.
\newblock Probabilistic matrix factorization.
\newblock In {\em NIPS}, 2007.

\bibitem[\protect\citeauthoryear{Pan \bgroup \em et al.\egroup }{2008}]{OneCF}
Rong Pan, Yunhong Zhou, Bin Cao, Nathan~Nan Liu, Rajan Lukose, Martin Scholz,
  and Qiang Yang.
\newblock One-class collaborative filtering.
\newblock In {\em ICDM}, 2008.

\bibitem[\protect\citeauthoryear{Pazzani}{1999}]{CFCBFdemograph99}
Michael~J Pazzani.
\newblock A framework for collaborative, content-based and demographic
  filtering.
\newblock {\em Artificial Intelligence Review}, 13(5-6):393--408, 1999.

\bibitem[\protect\citeauthoryear{Purushotham \bgroup \em et al.\egroup
  }{2012}]{CTRSoRec}
Sanjay Purushotham, Yan Liu, and C-c~J Kuo.
\newblock Collaborative topic regression with social matrix factorization for
  recommendation systems.
\newblock In {\em ICML}, 2012.

\bibitem[\protect\citeauthoryear{Sarwar \bgroup \em et al.\egroup
  }{2001}]{sarwar01:itemCF}
Badrul Sarwar, George Karypis, Joseph Konstan, and John Riedl.
\newblock Item-based collaborative filtering recommendation algorithms.
\newblock In {\em WWW}, 2001.

\bibitem[\protect\citeauthoryear{Tang and Liu}{2010}]{socialmedia}
Lei Tang and Huan Liu.
\newblock Community detection and mining in social media.
\newblock {\em Synthesis Lectures on Data Mining and Knowledge Discovery},
  2(1):1--137, 2010.

\bibitem[\protect\citeauthoryear{Tang \bgroup \em et al.\egroup
  }{2013}]{LOCABAL}
Jiliang Tang, Xia Hu, Huiji Gao, and Huan Liu.
\newblock Exploiting local and global social context for recommendation.
\newblock In {\em IJCAI}, 2013.

\bibitem[\protect\citeauthoryear{Wang and Blei}{2011}]{CTR}
Chong Wang and David~M Blei.
\newblock Collaborative topic modeling for recommending scientific articles.
\newblock In {\em KDD}, 2011.

\bibitem[\protect\citeauthoryear{Wang \bgroup \em et al.\egroup }{2014}]{dl4rs}
Hao Wang, Naiyan Wang, and Dit-Yan Yeung.
\newblock Collaborative deep learning for recommender systems.
\newblock {\em arXiv preprint arXiv:1409.2944}, 2014.

\bibitem[\protect\citeauthoryear{Zhang \bgroup \em et al.\egroup
  }{2013}]{localInfluence}
Jing Zhang, Biao Liu, Jie Tang, Ting Chen, and Juanzi Li.
\newblock Social influence locality for modeling retweeting behaviors.
\newblock In {\em IJCAI}, 2013.

\end{thebibliography}

\end{document}